\documentclass[amsmath,aps,prd,showpacs,a4paper,10pt]{revtex4}
\usepackage{epsf}
\usepackage{graphicx}  
\usepackage{dcolumn}   
\usepackage{bm}        

\begin{document}

\newcommand{\be}[1]{\begin{equation}\label{#1}}
\newcommand{\ee}{\end{equation}}
\newcommand{\bea}{\begin{eqnarray}}
\newcommand{\eea}{\end{eqnarray}}
\def\disp{\displaystyle}

\def\gsim{ \lower .75ex \hbox{$\sim$} \llap{\raise .27ex \hbox{$>$}} }
\def\lsim{ \lower .75ex \hbox{$\sim$} \llap{\raise .27ex \hbox{$<$}} }

\pagestyle{plain}

\begin{titlepage}

\begin{flushright}
hep-th/0412045
\end{flushright}

\vskip 1cm

\title{\Large \bf K-Chameleon and the Coincidence Problem}

\author{ Hao Wei\footnote{Email address: haowei@itp.ac.cn }}

\affiliation{ Institute of Theoretical Physics, Chinese Academy of
Sciences, P.O. Box 2735, Beijing 100080, China \\
 Graduate School of the Chinese Academy of Sciences, Beijing 100039, China}

\author{Rong-Gen Cai\footnote{Email address: cairg@itp.ac.cn}}
\affiliation{Institute of Theoretical Physics, Chinese Academy of
Sciences, P.O. Box 2735, Beijing 100080, China \\
CASPER, Department of Physics, Baylor University, Waco,
 TX76798-7316, USA}

\begin{abstract}
\centerline{\bf Abstract}

In this paper we present a hybrid model of k-essence and
chameleon, named as k-chameleon. In this model, due to the
chameleon mechanism, the directly strong coupling between the
k-chameleon field and matters (cold dark matters and baryons) is
allowed. In the radiation dominated epoch, the interaction between
the k-chameleon field and background matters can be neglected, the
behavior of the k-chameleon therefore is the same as that of the
ordinary k-essence. After the onset of matter domination, the
strong coupling between the k-chameleon and matters dramatically
changes the result of the ordinary k-essence. We find that during
the matter-dominated epoch, only two kinds of attractors may
exist: one is the familiar {\bf K} attractor and the other is a
completely {\em new}, dubbed {\bf C} attractor. Once the universe
is attracted into the {\bf C} attractor, the fraction energy
densities of the k-chameleon $\Omega_{\phi}$ and dust matter
$\Omega_m$ are fixed and comparable, and the universe will undergo
a power-law accelerated expansion. One can adjust the model so
that the {\bf K} attractor do not appear. Thus, the k-chameleon
model provides a natural solution to the cosmological coincidence
problem.

\end{abstract}

\pacs{98.80.-k, 98.80.Cq, 04.80.Cc}

\maketitle

\end{titlepage}

\newpage

\renewcommand{\baselinestretch}{1.5}

\setcounter{page}{1}


\section{Introduction}
There are now a lot of cosmological observations, such as SNe
Ia~\cite{r1,r2,r3}, WMAP~\cite{r4}, SDSS~\cite{r5} etc.. All
suggest that the universe is spatially flat, and consists of
approximately $70\%$ dark energy with negative pressure, $30\%$
dust matter (cold dark matter plus baryon), and negligible
radiation, and that the universe is undergoing an accelerated
expansion. To understand the nature of the dark energy remains as
one of biggest challenges to theorists and cosmologists~\cite{r6}.
The simplest candidate of the dark energy is a tiny positive
cosmological constant. However, it is difficult to understand why
the cosmological constant is about 120 orders of magnitude smaller
than its natural expectation, namely the Planck energy density.
This is the so-called cosmological constant problem. Another
puzzle of the dark energy is the cosmological coincidence problem,
i.e. why are the dark energy density and the dust matter energy
density comparable {\em now}? and why does the universe begin the
accelerated expansion just only {\em near recent}?

In order to have an interpretation to the accelerated expansion of
the universe, many alternatives to the cosmological constant have
been proposed. One of interesting scenarios is the so-called
quintessence model~\cite{r7}. The quintessence is a slowly varying
scalar field with a canonical kinetic energy term. With the
evolution of the universe, the scalar field slowly rolls down its
potential. A class of tracker solutions of
quintessence~\cite{r8,r9} is found in order to solve the
cosmological coincidence problem. As is shown, however, the
quintessence model still needs some fine-tuning in order for the
quintessence component to overtake the matter density at the
present epoch (for example, see Refs.~\cite{r12,r9}). Motivated by
the k-inflation~\cite{r10}, in which a scalar field with
non-canonical kinetic energy terms acts as the inflaton, the
so-called k-essence~\cite{r11,r12,r13,r14} is introduced to the
coincidence problem. In this well-known model, by the help of
non-linear kinetic energy terms, a dynamical solution to the
cosmological coincidence problem without fine-tuning is
possible~\cite{r11,r12}. In fact, k-essence is based on the idea
of a dynamical attractor solution which makes it act as a
cosmological constant only at the onset of matter domination.
Consequently, k-essence overtakes the matter energy density and
makes the universe start with accelerated expansion just near
recently. It is worth noting that to achieve a later-time
acceleration attractor, one needs to design the Lagrangian of the
model so that $r^{2}(y_d)>1$ in order to avoid the dust
attractor~\cite{r12}. After all, the quintessence and k-essence
fields are very light scalar fields. Such light fields may mediate
a long-range force, and therefore are subject to tight constraints
from the searches of the fifth force~\cite{r15} and the tests of
the equivalence principle (EP)~\cite{r16}.

On the other hand, the coupling between scalar field and matters
has been studied for some years (for example, see
Refs.~\cite{r17,r18,r19,r25,r26,r27,r28}). Recently, a novel
scenario named chameleon~\cite{r20,r21,r22,r23,r24} has been
proposed (see also \cite{Add}). In this scenario, the scalar field
can be directly
 coupled to matters (cold dark matters or baryons) with gravitational strength,
 in harmony with general expects from string theory, while this strong coupling
 can escape from the local tests of EP violations
 and fifth force searches. The basic idea of the chameleon scenario is that the
 scalar field acquires a mass which depends on the ambient matter density (so the name
chameleon). While nearly massless in the cosmos where the matter
density is tiny, the chameleon  mass is of order of an inverse
millimeter on the earth where the matter density is high, which is
sufficient to evade the tightest constraints from the tests of EP
violations and fifth force searches.

It is interesting to wonder what will happen when the k-essence is
 strongly coupled to matters through the chameleon mechanism. May
the virtues of k-essence and chameleon join together and the
shortcomings be avoided? The answer is yes. In this paper, we will
combine the k-essence with chameleon and present a so-called
 k-chameleon model, which of course is a
 hybrid of the k-essence and chameleon. Through the chameleon mechanism, the directly strong
 coupling between k-chameleon and matters (cold dark matters and baryons)
is allowed. We study the cosmological  evolution of k-chameleon
 and find that the k-chameleon model can provide a natural solution
to the cosmological coincidence problem.

In the k-chameleon model, during the radiation dominated epoch,
the interaction between the k-chameleon field and ambient matters
can be negligible. Therefore the behavior of the k-chameleon is
the completely same as that of the ordinary k-essence without the
interaction between the scalar field and background matters. As a
result, three kinds of attractors, namely {\bf R}, {\bf K}, {\bf
S} (following the notations of the k-essence model~\cite{r11,r12})
may exist. The radiation tracker, i.e. the {\bf R} attractor, has
the largest basin of attraction on the whole phase plane so that
most initial conditions join onto it and then makes this scenario
became insensitive to initial conditions. However, after the onset
of matter domination, the strong coupling between k-chameleon and
matters dramatically changes the result for the ordinary
k-essence. In the matter-dominated epoch, the {\bf D} and {\bf S}
attractors (which may exist in the ordinary k-essence model) are
physically forbidden due to the strong coupling between
k-chameleon and matters.  Note that unlike the ordinary k-essence
model, the disappearance of {\bf D} and {\bf S} attractors
naturally occurs in the k-chameleon model, need not any artificial
design of the Lagrangian. Actually, during the matter dominated
epoch, only two kinds of attractors may exist: one is the familiar
{\bf K} attractor and the other is a completely {\em new} one
named as {\bf C} attractor. The new attractor {\bf C} has some
desirable features which may provide a promising solution to the
cosmological coincidence problem.

Once the universe is attracted into the {\bf C} attractor, the
fraction energy densities of the k-chameleon $\Omega_{\phi}$ and
the matters $\Omega_m$ are fixed and they are comparable. Further,
many parameters, such as the parameter of equation-of-state of the
k-chameleon field $w_{\phi}$ and its kinetic energy term $X$, are
also fixed. And the universe will undergo a power-law accelerated
expansion forever. In this sense the k-chameleon model gives a
natural solution to the  cosmological coincidence problem. On the
other hand, note that if the kinetic energy term $X$ of the
k-chameleon is fixed at a somewhat small value (equivalently
$y\equiv 1/\sqrt{X}$ is large), the k-chameleon can be treated as
a canonical chameleon approximately. Therefore, we cannot detect
it from the tests of EP violation and fifth force searches on the
earth and in the solar system today, although it is strongly
coupled to background  matters.

This paper is organized as follows: In Sec.~\ref{sect2}, a brief
review of the chameleon mechanism is given. In Sec.~\ref{sect3},
we present our k-chameleon model and illustrate how the directly
strong coupling between k-chameleon and matters (cold dark matters
and baryons) is allowed while it cannot be detected from the tests
of EP violation and fifth force searches on the earth and in the
solar system. In Sec.~\ref{sect4}, the cosmological evolution of
k-chameleon is studied and the result shows that the k-chameleon
model may provide a promising solution to the cosmological
coincidence problem. A brief conclusion will be given in
Sec.~\ref{sect5}.

We use the units $\hbar=c=1$ throughout this paper. $M_{pl}\equiv
(8\pi G)^{-1/2}$ is the reduced Planck mass. We adopt the metric
convention as $(+\! -\! -\, -)$.


\section{\label{sect2} A BRIEF REVIEW OF THE CHAMELEON MECHANISM}
Following Refs.~\cite{r20,r21,r22}, consider a canonical chameleon
scalar field $\phi$ governed by the action
\be{eq1}
S=\int d^4
x\sqrt{-g}\left[-\frac{M_{pl}^2}{2}{\cal R}+\frac{1}{2}
g^{\mu\nu}\partial_{\mu}\phi\partial_{\nu}\phi-V(\phi)\right]+\int
d^4 x {\cal L}_{m}(\psi_{m}^{(i)}, g_{\mu\nu}^{(i)}), \ee where
$g$ is the determinant of the metric $g_{\mu\nu}$, ${\cal R}$ is
the
 Ricci scalar and $\psi_{m}^{(i)}$ are various matter fields labeled by $i$.
 In the chameleon mechanism, the scalar field $\phi$ is supposed to
 directly interact with matters through a conformal coupling. In other words,
 each matter field $\psi_{m}^{(i)}$ couples to a metric $g_{\mu\nu}^{(i)}$ which
 is related to the Einstein-frame metric $g_{\mu\nu}$ by the rescaling
 \be{eq2}
g_{\mu\nu}^{(i)}=e^{2\beta_{i}\phi/M_{pl}}g_{\mu\nu}, \ee where
$\beta_{i}$ are dimensionless constants. Moreover, the different
$\psi_{m}^{(i)}$ fields are assumed  not to interact with each
other for simplicity. From the action Eq.~(\ref{eq1}), the
equation of motion for $\phi$ is
\be{eq3}
\nabla^{2}\phi=-V_{,\phi}-\sum\limits_{i}\frac{\beta_i}{M_{pl}}
e^{4\beta_{i}\phi/M_{pl}}g^{\mu\nu}_{(i)}T_{\mu\nu}^{(i)},
\ee
where
$$\nabla^2\phi\equiv g^{\mu\nu}\nabla_{\mu}\nabla_{\nu}\phi
=\frac{1}{\sqrt{-g}}\partial_{\mu}\left[\sqrt{-g}g^{\mu\nu}
\partial_{\nu}\phi\right]$$
and $T_{\mu\nu}^{(i)}=(2/\sqrt{-g^{(i)}})\delta {\cal L}_{m}/\delta g^{\mu\nu}_{(i)}$
is the stress-energy tensor density for the $i$-th form of matter. $V_{,\phi}$
 denotes the derivative of $V$ with respect to $\phi$. For non-relativistic
  dust-like matter, $g^{\mu\nu}_{(i)}T_{\mu\nu}^{(i)}=\tilde{\rho}_{i}$,
  where $\tilde{\rho}_{i}$ is the energy density. Defined in this way,
  however, $\tilde{\rho}_{i}$ is not conserved in Einstein frame. Instead,
   it is more convenient to define a matter density
   $\rho_{mi}\equiv\tilde{\rho}_{i}\, e^{3\beta_{i}\phi/M_{pl}}$
   which is independent of $\phi$ and is conserved in Einstein frame.
   Thus, Eq.~(\ref{eq3}) can be recast as
\be{eq4}
\nabla^{2}\phi=-V_{,\phi}-\sum\limits_{i}\frac{\beta_i}{M_{pl}}\rho_{mi}\,
 e^{\beta_{i}\phi/M_{pl}}=-V^{eff}_{,\phi}.
\ee
 Note that the dynamics of $\phi$ is not governed solely by
$V(\phi)$, but rather by an effective potential
  \be{eq5}
V_{eff}(\phi)=V(\phi)+\sum\limits_{i}\rho_{mi}\,
e^{\beta_{i}\phi/M_{pl}},
 \ee
  which depends explicitly on the matter density $\rho_{mi}$. The key
  ingredient to achieve a
successful chameleon model is that the effective potential
$V_{eff}(\phi)$ has a minimum even when $V(\phi)$ is monotonic. In
fact, if $V(\phi)$ is monotonically decreasing and $\beta_i>0$ or,
equivalently, $V(\phi)$ is monotonically increasing and
$\beta_i<0$, the effective potential $V_{eff}(\phi)$ has a minimum
$\phi_{min}$ satisfying
\be{eq6}
V^{eff}_{,\phi}(\phi_{min})=V_{,\phi}(\phi_{min})+\sum\limits_{i}
\frac{\beta_i}{M_{pl}}\rho_{mi}\,
e^{\beta_{i}\phi_{min}/M_{pl}}=0.
 \ee
  Meanwhile, the mass of small
fluctuations about the minimum $\phi_{min}$ is
\be{eq7}
m_{eff}^2\equiv V^{eff}_{,\phi\phi}(\phi_{min})
=V_{,\phi\phi}(\phi_{min})+\sum\limits_{i}\frac{\beta_{i}^{2}}{M_{pl}^{2}}\rho_{mi}\,
e^{\beta_{i}\phi_{min}/M_{pl}}.
 \ee
  In other words, the originally
massless scalar field acquires a mass which depends on the local
matter density. The denser the environment, the more massive the
chameleon is. Actually, while the coupling constants $\beta_i$ can
be of order unity as the natural expectations from string theory,
it is still possible for the mass of chameleon, i.e. $m_{eff}$, to
be sufficiently large on the earth to evade current constraints on
EP violation and fifth force. On the other hand, through the
so-called ``thin-shell'' effect, the chameleon-mediated force
between two large objects, such as the earth and the sun, is much
suppressed, which thereby ensures that solar system tests of
gravity are satisfied. For more details, see the original
papers~\cite{r20,r21,r22,r23}. The quantum stability analysis of
the chameleon model is presented in Ref.~\cite{r24}.

It is worth noting that, in most existing canonical chameleon
models, the potentials $V(\phi)$ are assumed to be of the runaway
form, namely it is monotonically decreasing and satisfies
$$\lim\limits_{\phi\to\infty}V=0,
~~~~~~~\lim\limits_{\phi\to\infty}\frac{V_{,\phi}}{V}=0,
~~~~~~~\lim\limits_{\phi\to\infty}\frac{V_{,\phi\phi}}{V_{,\phi}}=0
~~~\dots$$
as well as
$$\lim\limits_{\phi\to 0}V=\infty,
~~~~~~~\lim\limits_{\phi\to 0}\frac{V_{,\phi}}{V}=\infty,
~~~~~~~\lim\limits_{\phi\to
0}\frac{V_{,\phi\phi}}{V_{,\phi}}=\infty~~~\dots.$$
 Actually, the fiducial potentials are chosen to be
 \be{eq8}
V(\phi)=M^{4}\left(\frac{M}{\phi}\right)^n~~~~~~~{\rm
and}~~~~~~~V(\phi)=M^{4}\exp(M^{n}/\phi^{n})
 \ee
  in Ref.~\cite{r20,r22} and Ref.~\cite{r21}, respectively. However,
for the potentials given in Eq.~(\ref{eq8}), to achieve a
successful chameleon model,  the mass scale $M$ has to statisfy
 \be{eq9}
M\,\lsim\, 10^{-3}\,{\rm eV}, \ee
 which is about 30
orders of magnitude smaller than its natural expectation, namely
the Planck mass.  Fortunately,  Ref.~\cite{r23} shows that  a
chameleon model with a non-runaway form potential \be{eq10}
V(\phi)=\frac{1}{2}\, m_{\phi}^2\,\phi^2+\frac{\xi}{4!}\,\phi^4
\ee
 can be a successful example even when the parameter $\xi$
is of order unity.  In Sec.~\ref{newchameleon} of the present
paper, a completely {\em new} canonical chameleon with non-runaway
potential and without any fine-tuning like in Eq.~(\ref{eq9}) will
be presented. This new canonical chameleon model is another
successful example.


\section{\label{sect3} K-CHAMELEON MODEL}

So far, all chameleon models existing in the literature  is of the
form of quintessence-like, namely the kinetic energy term of the
scalar field is canonical one. As is well-known, non-linear
kinetic energy terms naturally appear in many models unifying
gravity with other particle forces, including supergravity and
superstring theory. For many years, the contributions  of these
higher order terms have been ignored for the reasons of
simplicity. The example of k-essence~\cite{r11,r12,r13,r14}
demonstrates that the effects of non-linear dynamics can be
dramatic. Here, motivated by k-essence, we put the chameleon and
k-essence together and present a k-chameleon model.  Namely,
 we consider a scalar field with non-linear kinetic
terms and the scalar field is strongly coupled to matters.

\subsection{Setup}
Our starting point is the action
\be{eq11}
S=\int d^4 x\sqrt{-g}\left[-\frac{M_{pl}^2}{2}{\cal R}
+p(\phi, X)\right]+\int d^4 x {\cal L}_{m}(\psi_{m}^{(i)},
 g_{\mu\nu}^{(i)}),
\ee
where $g$ is the determinant of the metric $g_{\mu\nu}$,
 ${\cal R}$ is the Ricci scalar and $\psi_{m}^{(i)}$ are
 various matter fields labeled by $i$. The scalar field $\phi$
 interacts directly with matters through a conformal coupling.
 Explicitly, each matter field $\psi_{m}^{(i)}$ couples to a metric
  $g_{\mu\nu}^{(i)}$ which is related to the Einstein-frame metric
  $g_{\mu\nu}$ by the rescaling
\be{eq12}
 g_{\mu\nu}^{(i)}=A_{i}^{2}(\phi)g_{\mu\nu},
\ee
 where $A_i(\phi)$ is a function of the k-chameleon field $\phi$.
   Note that Eqs.~(\ref{eq11}) and~(\ref{eq12}) are of the
general form arising from string theory, supergravity and
Brans-Dicke theory. Moreover, the different fields
$\psi_{m}^{(i)}$ are assumed not to interact with each other for
simplicity. The kinetic energy term is defined by \be{eq13}
X\equiv\frac{1}{2}g^{\mu\nu}\partial_{\mu}\phi\partial_{\nu}\phi
\, .
 \ee
Note that if one takes $p\, (\phi,X)=X-V(\phi)$ and
$A_{i}(\phi)=\exp(\beta_{i}\phi/M_{pl})$, this case then reduces
to the canonical chameleon model presented in Sec.~\ref{sect2}.

Varying the action Eq.~(\ref{eq11}) with respect to $\phi$ yields
the equation of motion for the k-chameleon field $\phi$
\be{eq14}
\frac{1}{\sqrt{-g}}\partial_{\mu}\left[\sqrt{-g}\, p_{,_X}\,
g^{\mu\nu}\partial_{\nu}\phi\right]=p_{,_\phi}-\sum\limits_{i}\alpha_{i}(\phi)\,
A_{i}^{4}(\phi)\, g^{\mu\nu}_{(i)}\, T_{\mu\nu}^{(i)},
 \ee
  where $p_{,_X}$ denotes the derivative of $p$ with respect to $X$, and
\be{eq15}
T_{\mu\nu}^{(i)}=\frac{2}{\sqrt{-g^{(i)}}}\frac{\delta
{\cal L}_{m}}{\delta g^{\mu\nu}_{(i)}}
 \ee
 is the stress-energy tensor density for the $i$-th form of matter,
 and
  \be{eq16}
 \alpha_{i}(\phi)\equiv\frac{\partial\ln
 A_{i}(\phi)}{\partial\phi}.
 \ee
  For the $i$-th non-relativistic
dust-like matter, the energy density in Einstein frame
\be{eq17}
\rho_i=T^{\mu}_{~\mu}=g^{\mu\nu}\frac{2}{\sqrt{-g}}\frac{\delta{\cal
L}_m}{\delta g^{\mu\nu}}=A_{i}^{4}(\phi)\tilde{\rho}_i,
 \ee
  where
  \be{eq18}
   \tilde{\rho}_i=g^{\mu\nu}_{(i)}\,
T_{\mu\nu}^{(i)}=g^{\mu\nu}_{(i)}\frac{2}{\sqrt{-g^{(i)}}}\frac{\delta{\cal
L}_m}{\delta g^{\mu\nu}_{(i)}}
 \ee
  is the energy density in the
matter frame. However, $\tilde{\rho}_{i}$ is not conserved in
Einstein frame. Instead, it is more convenient to define
 matter density
  \be{eq19}
 \rho_{mi}\equiv A_{i}^{3}(\phi)\tilde{\rho}_{i}
 \ee
 which is independent of $\phi$ and conserved in Einstein frame.
  Then Eq.~{(\ref{eq14})} can be rewritten as
   \be{eq20}
 \frac{1}{\sqrt{-g}}\partial_{\mu}\left[\sqrt{-g}\, p_{,_X}\,
g^{\mu\nu}\partial_{\nu}\phi\right]=p_{,_\phi}-\sum\limits_{i}
 \alpha_{i}(\phi)\,\rho_{i}.
\ee
 In addition, form the action Eq.~(\ref{eq11}), we have the pressure
and energy density of the k-chameleon field $\phi$~\cite{r10}
 \be{eq21}
p_\phi=p(\phi,X),~~~~~~~\rho_\phi=2Xp_{,_X}-p\ ,
 \ee
 respectively.
 Consider a flat FRW universe, whose metric is
 \be{eq22}
ds^2=dt^2-a^2(t)d{\bf x}^2,
 \ee
 where $a$ is the scale factor. If the scalar field $\phi$ is spatially homogeneous,
 one then has
 \be{eq23}
X=\frac{1}{2}\dot{\phi}^2,
 \ee
 where a dot denotes the derivative
with respect to the cosmic time $t$. Furthermore, by using
Eq.~{(\ref{eq21})}, the equation of motion for the k-chameleon
field $\phi$, namely Eq.~{(\ref{eq20})}, can be recast as
\be{eq24}
\dot{\rho}_{\phi}+3H(\rho_\phi+p_\phi)=-\sum\limits_{i}\alpha_{i}
(\phi)\,\rho_{i}\,\dot{\phi}, \ee where $H=\dot{a}/a$ is the
Hubble parameter.
 From the total energy conservation equation
\be{eq25}
\dot{\rho}_{tot}+3H(\rho_{tot}+p_{tot})=0,
 \ee
   where the total pressure and energy density are
 \be{eq26}
p_{tot}=p_\phi+p_r,~~~~~{\rm and}
~~~~~\rho_{tot}=\rho_\phi+\sum\limits_{i}\rho_i+\rho_r, \ee
respectively, we have
  \be{eq27}
\dot{\rho}_{i}+3H\rho_i=\alpha_{i}(\phi)\,\rho_{i}\,\dot{\phi},
\ee and \be{eq28} \dot{\rho}_{r}+4H\rho_{r}=0,
 \ee
  where $p_r=\rho_r/3$ and $\rho_r$ are the pressure and energy density of
radiation respectively. Note that there is no coupling between the
scalar field $\phi$ and radiation because the trace of the
stress-energy tensor of radiation vanishes. Finally we write down
the Friedmann equation
  \be{eq29}
3H^{2}M_{pl}^2=\rho_{tot}=\rho_\phi+\rho_m+\rho_r
 \ee
where
$\rho_m=\sum\limits_{i}\rho_i$ is the sum of the energy density of
all matter components.

\subsection{Master equations of k-chameleon}

In this paper, following k-essence~\cite{r11,r12}, we only
consider a  factorizable Lagrangian of the form
 \be{eq30}
p\,(\phi,X)=K(\phi)D(X),
\ee
 where we assume  $K(\phi)>0$.
From Eq.~{(\ref{eq21})}, we have
 \bea
&p_\phi=p\,(\phi,X)=K(\phi)D(X),\nonumber\\
&\rho_\phi=2Xp_{,_X}-p=K(\phi)\,[\, 2XD_{,_X}(X)-D(X)]\equiv K(\phi)E(X). \label{eq31}
 \eea
The parameter of equation-of-state is \be{eq32}
w_\phi\equiv\frac{p_\phi}{\rho_\phi}=\frac{D}{E}=\frac{D}{2XD_{,_X}-D}.
\ee
 According to the definition in~\cite{r10}, the sound speed is
  \be{eq33}
C_{s}^{2}=\frac{p_{\phi,_X}}{\rho_{\phi,_X}}=\frac{D_{,_X}}{E_{,_X}}.
\ee
 Substituting Eqs.~{(\ref{eq31}) and (\ref{eq23})} into Eq.~{(\ref{eq24})},
 we  obtain
  \be{eq34}
\frac{dX}{dN}=-\frac{E}{E_{,_X}}\left[3(1+w_\phi)+\sigma\frac{K_{,\phi}}{K}
\frac{\sqrt{2X}}{H}+\sigma\sum\limits_{i}\alpha_{i}(\phi)
\rho_{i}\frac{\sqrt{2X}}{HKE}\right],
\ee
where $N\equiv\ln a$ is the so-called e-folding time, $\sigma$
is the sign of $\dot{\phi}$.

It is convenient to re-express $D$ as $D=g(y)/y$ and to view it as
a function of the new variable $y\equiv 1/\sqrt{X}$.
Eqs.~{(\ref{eq30})--(\ref{eq33})} then become
\bea
&p_\phi=\disp \frac{Kg}{y},~~~~~~~\rho_\phi=\disp -Kg^{\prime}, \label{eq35}\\
&w_\phi=\disp
-\frac{g}{g^{\prime}y},~~~~~~~C_{s}^2=\disp\frac{g-g^{\prime}y}
{g^{\prime\prime}y^2}, \label{eq36}
 \eea
   where a prime denotes the
derivative with respect to $y$. Taking into account
Eqs.~{(\ref{eq29})}, (\ref{eq35}) and (\ref{eq36}), one can recast
Eq.~{(\ref{eq34})} in terms of the new variable
 as \be{eq37}
\frac{dy}{dN}=\frac{3}{2}\,\frac{w_{\phi}(y)-1}{s^{\prime}(y)}
\left[s(y)+\sigma\frac{K_{,\phi}}{2K^{3/2}}\sqrt{\frac{\rho_\phi}
{\rho_{tot}}}+\sigma\sum\limits_{i}\frac{\alpha_{i}(\phi)}{2K^{1/2}}
\frac{\rho_i}{\sqrt{\rho_{\phi}\rho_{tot}}}\right],
\ee
 where
  \be{eq38}
s(y)=\left(-\frac{3g^{\prime}}{8M_{pl}^{2}}\right)^{1/2}y(1+w_\phi)
=\sqrt{\frac{3}{8M_{pl}^{2}}}\,\frac{g-g^{\prime}y}{\sqrt{-g^{\prime}}}.
\ee
 Note that the requirements of positivity of the energy
density, $\rho_\phi>0$, and stability of the k-chameleon
background, $C_{s}^{2}>0$, imply
 \be{eq39}
-g^{\prime}>0,~~~~~~~~~~g^{\prime\prime}>0.
 \ee
  These conditions indicate that $g$ should be a decreasing convex function of
$y=1/\sqrt{X}$. A sample of the function $g(y)$ is plotted in
Fig.~1.

Since we are attempting to study the cosmological evolution of
k-chameleon and try to find out its attractors, we impose the
condition that the coefficients of the last two terms in
Eq.~{(\ref{eq37})} to be constants for simplicity. Thus,
 $K(\phi)$, $\alpha_{i}(\phi)$ and $A_{i}(\phi)$ should be the form
\be{eq40}
K(\phi)=\frac{M^2}{\phi^2},~~~~~~~~~~\alpha_{i}(\phi)=\frac{\beta_i}{\phi},
~~~~~~~~~~A_{i}(\phi)=\left(\frac{\phi}{M}\right)^{\beta_i}, \ee
where $M$ is a constant mass scale and $\beta_i$  are
dimensionless positive constants.
 Furthermore, although $\beta_i$ may be different for different matter species,
 we take a same value $\beta$ for all $\beta_i$ for simplicity [$\alpha(\phi)$,
 $A(\phi)$ for all $\alpha_{i}(\phi)$, $A_{i}(\phi)$ accordingly]. In fact,
 it is straightforward to generalize it to the generic case with different $\beta_i$.
 In addition, without
 loss of generality, we restrict ourselves to the most interesting case of
 positive $\dot{\phi}$, namely $\sigma=+1$. Under these simplifications,
  Eq.~{(\ref{eq37})} becomes
\be{eq41}
\frac{dy}{dN}=\frac{3}{2}\,\frac{w_{\phi}(y)-1}{r^{\prime}(y)}
\left[r(y)-\sqrt{\Omega_\phi}+\frac{\beta}{2}\frac{\Omega_m}
{\sqrt{\Omega_\phi}}\right],
\ee
 where $r(y)\equiv Ms(y)$ is a
dimensionless function of $y$. It is worth noting that if
$M=\sqrt{3}M_{pl}$,  $r(y)$ then is completely the same as that of
the k-essence case~\cite{r11,r12}. The fraction energy densities
$\Omega_\phi\equiv\rho_\phi/\rho_{tot}$ and
$\Omega_m\equiv\rho_m/\rho_{tot}$, where
$\rho_m=\sum\limits_{i}\rho_i$ is the sum of the energy density of
all matter components. At the same time, from
Eqs.~{(\ref{eq24})--(\ref{eq28})}, we reach
 \be{eq42}
\frac{d\Omega_\phi}{dN}=-\frac{\alpha}{H}\,\Omega_{m}\dot{\phi}+3\,
\Omega_{\phi}(w_{tot}-w_\phi),
\ee
\be{eq43}
\frac{d\Omega_m}{dN}=\frac{\alpha}{H}\,\Omega_{m}\dot{\phi}+3\,\Omega_{m}w_{tot},
\ee
where
 \be{eq44}
w_{tot}\equiv\frac{p_{tot}}{\rho_{tot}}=w_{\phi}\Omega_\phi+\frac{1}{3}\,\Omega_r,
\ee
 and $\Omega_r\equiv\rho_r/\rho_{tot}$ is the fraction
energy density of radiation.

Thus we have obtained the master equations governing the whole
system. Compared to the ordinary k-essence model~\cite{r11,r12},
we find that some new terms which describe the coupling between
k-chameleon and matters enter these equations. In addition, let us
mention that in our k-chameleon model, the coupling function
$A(\phi)$ is of a form of power-law [see Eq.~(\ref{eq40})], while
in the ordinary chameleon models~\cite{r20,r21,r22,r23,r24} or
other coupled models of scalar field to matters in the
literature~\cite{r18,r19}, the coupling function is of either the
exponential type~\cite{r18,r20,r21,r22,r23,r24}
 or linear type~\cite{r19}.


\subsection{\label{newchameleon} Coupling constant $\beta$ and  mass scale $M$}

In this subsection we will discuss the constraints on the coupling
constant $\beta$ and the mass scale $M$ imposed by the fifth force
experiment~\cite{r15} and the tests of the equivalence principle
(EP)~\cite{r16}, in order to obtain a successful chameleon
mechanism.

Note that all the searches of the fifth force and EP violation
have been performed only at the present epoch  on the earth or in
the solar system.  This point is very important. As we will see,
in the k-chameleon model,  the kinetic energy term $X$ of the
k-chameleon field could be fixed at a somewhat small value
(equivalently $y\equiv 1/\sqrt{X}$ is large) at the current
accelerated expansion epoch. In this case, the non-linear kinetic
energy terms can be neglected and the k-chameleon can be treated
as a canonical chameleon approximately.

In this case, $D(X)$ in the Lagrangian $p(\phi,X)=K(\phi)D(X)$ can
be approximated to
\be{eq45}
 D(X)\simeq c_1 X-c_2,
  \ee where $c_1$
is a dimensionless positive constant and $c_2$ is a positive
constant with dimension of energy density. To change the
Lagrangian to a canonical form, we make a redefinition of the
field variable.  Introducing a new scalar field
\be{eq46}
\phi_{new}=-\sqrt{c_1}M\ln\frac{\phi}{M}
\ee satisfying
$$X_{new}=\frac{1}{2}g^{\mu\nu}\partial_{\mu}\phi_{new}
\partial_{\nu}\phi_{new}=c_{1}K(\phi)X,$$
where
  $K(\phi)=M^{2}/\phi^2$ has been considered. Accordingly, the
conformal coupling $A(\phi)=(\phi/M)^\beta$ is transformed to
\be{eq47}
A(\phi_{new})=\exp\left[-\frac{\beta\phi_{new}}{\sqrt{c_1}M}\right],
\ee
 and the Lagrangian becomes the canonical one,
  \be{eq48}
p(\phi_{new},X_{new})=X_{new}-V(\phi_{new}),
 \ee
  where the potential
  \be{eq49}
V(\phi_{new})=c_2\exp\left[\frac{2\phi_{new}}{\sqrt{c_1}M}\right].
  \ee
   The equation of motion Eq.~{(\ref{eq20})} then becomes
\be{eq50}
\nabla^{2}\phi_{new}=-V^{eff}_{,\phi_{new}}(\phi_{new}),
 \ee
  where the effective potential
  \be{eq51}
V_{eff}(\phi_{new})=V(\phi_{new})+A(\phi_{new})\rho_{mt}^{new},
 \ee
and the matter energy density
   \be{eq52}
\rho_{mt}^{new}=A^{3}(\phi_{new})\sum\limits_{i}\tilde{\rho}_{i},
 \ee
which is independent of $\phi_{new}$ and is conserved in Einstein
frame, and $\tilde{\rho}_{i}$ is defined in Eq.~{(\ref{eq18})}.
Obviously,  because $V^{eff}_{,\phi_{new}\phi_{new}}(\phi_{new})$
is always larger than zero,  the effective potential has a minimum
at
\be{eq53}
\phi_{new}^{min}=\frac{\sqrt{c_1}M}{2+\beta}\ln\frac{\beta\rho_{mt}^{new}}{2c_2}
\ee
satisfying $V^{eff}_{,\phi_{new}}(\phi_{new}^{min})=0$. Thus,
the mass of small fluctuations about the minimum
$\phi_{new}^{min}$ is
\be{eq54}
 m^{2}_{new}\equiv
V^{eff}_{,\phi_{new}\phi_{new}}(\phi_{new}^{min}).
\ee
 After some algebra, we get
  \be{eq55}
m_{new}^{-1}=\left(\frac{c_1}{\beta^2+2\beta}\right)^{1/2}
\left(\frac{M}{M_{pl}}\right)\left(\frac{\beta
M_{pl}^4}{2c_2}\right)^{\beta/(4+2\beta)}\left(\frac{\rho_{mt}^{new}}
{M_{pl}^4}\right)^{-1/(2+\beta)}M_{pl}^{-1}.
\ee
 It is easy to see that, if one takes the natural expectations
of the constants as
 \be{eq56}
  c_1\sim\beta\sim{\cal
O}(1),~~~~~~~c_2\sim{\cal O}(M_{pl}^4)~~~~~~~{\rm
and}~~~~~~~M\sim{\cal O}(M_{pl}),
\ee
  $m_{new}^{-1}$ is indeed a
small quantity for any reasonable matter density. For instance,
the atmosphere has mean density $\rho_{atm}\sim 10^{-3}~{\rm
g/cm^3}$. Substituting into Eq.~{(\ref{eq55})} and assuming
Eq.~{(\ref{eq56})}, we find
 \be{eq57}
  m_{atm}^{-1}\sim{\cal O}(1~{\rm mm}),
\ee
 which is sufficient~\cite{r20,r23} to evade
current constraints on EP violation and fifth force. Note that the
field redefinition Eq.~{(\ref{eq46})} can be taken as
$\phi_{new}=\sqrt{c_1}M\ln\, (\phi/M)$ and the final result
obtained above is still valid. In addition, let us stress here
that the potential Eq.~(\ref{eq49}) is different from those in
Refs.\cite{r20,r21,r22,r23}. Therefore it is interesting to
further study this model more details and other aspects such as,
the ``thin-shell'' effect. We will present these details in a
separate paper~\cite{r29}.

The upshot of this subsection is that we illustrate briefly how
the directly strong coupling between k-chameleon and matters (cold
dark matters and baryons) is allowed while we cannot detect it
from the tests of EP violations and fifth force searches on the
earth or in the solar system today. The coupling constant $\beta$
needs not to be tuned to an extremely small value and can be of
order unity, in harmony with general expectations from string
theory. The mass scale $M$ can be of order $M_{pl}$. In other
words, it needs not to be tuned like Eq.~{(\ref{eq9})} mentioned
above as in Refs.~\cite{r20,r21,r22}.


\section{\label{sect4} COSMOLOGICAL EVOLUTION OF K-CHAMELEON AND THE COSMOLOGICAL
 COINCIDENCE PROBLEM}

In this section, we will study the cosmological evolution of
k-chameleon. Just like the cases of quintessence~\cite{r8,r9} and
k-essence~\cite{r11,r12,r13,r14}, the key point is to find out its
attractors during the evolution. During the radiation- and
matter-dominated epochs, there are several sets of possible
attractors in the k-chameleon model.  Note that, as illustrated in
Sec.~\ref{newchameleon}, the mass scale $M$ can be of order
 $M_{pl}$. Thus, for simplicity, we take $M=\sqrt{3}M_{pl}$ from now on.


\subsection{\label{rd} Radiation-dominated epoch}

In this epoch, the fraction energy density of matter
$\Omega_m\simeq 0$. One can see that all terms describing the
coupling between k-chameleon and matters
 can be neglected, and the master equations governing the system, namely
  Eqs.~{(\ref{eq41})--(\ref{eq44})}, reduce to corresponding ones
  for the ordinary k-essence case~\cite{r11,r12}. Thus, the cosmological
  evolution of k-chameleon in this epoch is completely the same as that  of
  ordinary k-essence. A detailed study was presented in Ref.~\cite{r12} for this case.
  Therefore we will not repeat here and only mention some key points and
  make some remarks.

  \begin{figure}[ht]
\includegraphics[totalheight=3.5in,angle=90]{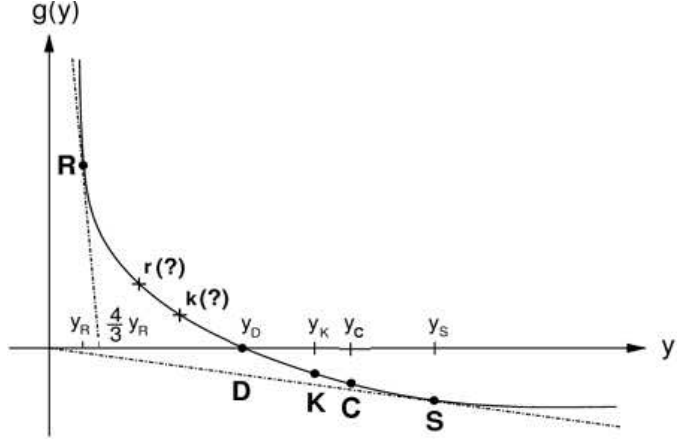}
 \caption{A sample of function $g(y)$, reproduced from \cite{r12}. Here the new
 attractor ${\bf C}$ appears between $y_d <y< y_s$.}
\end{figure}

\subsubsection{Attractors}

During the radiation dominated epoch, three kinds of attractors,
{\bf R} (radiation), {\bf K} (k-field), {\bf S} (de Sitter)
attractors (following notations of the k-essence
model~\cite{r11,r12}) may exist.
  The location of the {\bf R} attractor, i.e. $y_r=const.$, is determined by
\be{eq58}
y_{r}g^{\prime}(y_r)=-3g(y_r),
\ee
 which corresponds to
$y_r<y_d$ to ensure $g>0$ (positive pressure), where $y_d$ is the
location of the function $g(y)$ across the $y$ coordinate axis,
 i.e. $g(y_d)=0$ (see Fig.~1 of the present paper or Fig.~1 of Ref.~\cite{r12}). The fraction energy
 density of k-chameleon is given by
\be{eq59}
\Omega_{\phi}^{(r)}=r^{2}(y_r)=-2g^{\prime}(y_r)y_{r}^2,
\ee
and the {\bf R} attractor exists only if $r^{2}(y_r)<1$. The
parameter of equation-of-state $w_{\phi}(y_r)=1/3$, and the
k-chameleon mimics the radiation. Note that
$\Omega_{\phi}^{(r)}=r^{2}(y_r)$ has to be in the
 range $1-10\%$ in order to satisfy the constraints from the big bang
 nucleosynthesis (BBN)~\cite{r30,r31}.

 The location $y_k$ of the second
 attractor {\bf K} is determined by
\be{eq60} \Omega_{\phi}^{(k)}=r^{2}(y_k)=1,
 \ee
  which implies that the k-chameleon dominates over other components.
  The parameter of
equation-of-state
\be{eq61}
w_{\phi}(y_k)=-1+\frac{2\sqrt{2}}{3}\frac{1}{\sqrt{-g_{k}^{\prime}y_{k}^2}}=const.,
\ee
and the scale factor is
 \be{eq62}
 a\propto
t^{2/3[1+w_{\phi}(y_k)]}=t^{\sqrt{-g_{k}^{\prime}y_{k}^{2}/2}}.
\ee
  If $\sqrt{-g_{k}^{\prime}y_{k}^{2}/2}>1$, the solution describes
a power-law inflation. Physically, this condition is equivalent to
$w_{tot}(y_k)=w_{\phi}(y_k)<-1/3$.

The third attractor {\bf S} is defined by $w_{\phi}(y_s)=-1$ and
$\Omega_\phi^{(s)}=r^2(y_s) \simeq 0$.
 From Eq.~{(\ref{eq36})}, we have
  \be{eq63}
g(y_s)=g^{\prime}_{s}y_s,~~~~~~~C_{s}^2=0. \ee Geometrically, {\bf
S} is a fixed point on the curve $g(y)$, which makes the tangent
of $g$ at $y_s$ passes through the origin (see Fig.~1). It is
clear that $y_s$ always exists for the convex decreasing function
$g(y)$ unless $y_s\to\infty$.

The combination of cosmologically relevant attractors during the
 radiation-dominated epoch can be one of three types:
\begin{itemize}
\item {\bf R}, {\bf S} and {\em no} other attractors at $y_r<y<y_s$;
\item {\bf R}, {\bf S}, {\bf K} plus possibly other attractors at $y<y_d$;
\item {\bf R}, {\bf S} ({\em no} {\bf K} attractor) and at least one additional
attractor {\bf r}(?) or {\bf k}(?) (following notations of k-essence).
\end{itemize}
 The phase diagrams for these three cases have been drawn in Figs.~2,
3 and 4 of Ref.~\cite{r12}, respectively. Obviously, one can see
from these phase diagrams that the {\bf R} attractor has the
largest basin of attraction on the whole phase plane for all these
three cases.

\subsubsection{\label{remarks} Remarks}

The stability analysis of the attractors was done in
Ref.~\cite{r12}.  With a closer look, we find that the stability
analysis is valid only for the {\bf R} and {\bf D} (dust)
attractor (which may appear during the matter dominated epoch in
the ordinary k-essence model). Actually, because the {\bf K} and
{\bf S} are not determined by $w_\phi=w_m$ [following notations of
Ref.~\cite{r12}, $w_m$ denotes the state parameter of background
matter (radiation or dust)],
 Eq.~{(24)} of Ref.~\cite{r12} is not valid for them. In fact, Eq.~{(24)} of
 Ref.~\cite{r12} should be changed to $d\delta\Omega_{\phi}/dN=3w_{\phi}
 (y_k)\delta\Omega_{\phi}$
 and $d\delta\Omega_{\phi}/dN=3\delta\Omega_{\phi}$ for {\bf K} and {\bf S},
  respectively. Therefore, the stability condition for {\bf K} attractor
  is $\sqrt{-g_{k}^{\prime}y_{k}^{2}/2}>2/3$ while {\bf S} is always {\em unstable}.
This point can be seen clearly in the phase diagrams, i.e.
   Figs.~{2--6} of Ref.~\cite{r12}: the phase flow can get near to {\bf S}, but
   never reaches it, and the phase flow is then forced to leave it.

The second remark is about the basin of attraction of the {\bf R}
attractor. In Ref.~\cite{r14}, the statement ``{\bf R} attractor
has the largest basin of attraction on the whole phase plane in
the radiation-dominated epoch'' was criticized by numerically
analyzing two concrete models.  In these two models the function
$D(X)$ in the Lagrangian has the following forms
$$D(X)=-2.01+2\sqrt{1+X}+3\times10^{-17}X^3-10^{-24}X^4,$$
and
$$D(X)=-2.05+2\sqrt{1+f(X)},$$
where
$$f(X)=X-10^{-8}X^2+10^{-12}X^3-10^{-16}X^4+10^{-20}X^5-10^{-24}X^6/2^6,$$
which are first given in Refs.~\cite{r11} and \cite{r12},
respectively. (Note that in Refs.~\cite{r11,r12,r14}, the unit
$3M_{pl}^2=1$ was used.) However,  we notice that the conclusion
``{\bf R} attractor has the largest basin of attraction on the
whole phase plane in the radiation-dominated epoch'' is based on
another non-analyzable Lagrangian, i.e. Eq.~(42) of
Ref.~\cite{r12},
$$g(y)=g_{glue}(y)\left(1-\frac{y}{s^{2}y_d}\right),$$
where $g(y)$ is a function parametrized by five parameters:
 $y_r$, $g^{\prime}_r$, $y_d$, $g^{\prime}_d$ and $s^{2}y_d$
  (see Sec. V of Ref.~\cite{r12} for more detail). By choosing
  suitable parameters, as is shown in Sec. VA and VB of Ref.~\cite{r12},
  that the {\bf R} attractor has indeed the largest basin of attraction on the
  whole phase plane in the radiation-dominated epoch is possible.

The third remark we would like to stress is about the implication
of $y_s$ mentioned above. In the left side of $y_s$, i.e. $y<y_s$,
 one has $w_{\phi}>-1$ and $C_{s}^2>0$ while $g^{\prime\prime}>0$. Because
 the $g^{\prime\prime}$ is a continuous function, and the sound speed $C_{s}^2$
 cannot diverge at $y_s$, which implies that $g^{\prime\prime}\not=0$ at $y_s$.
 Therefore $g^{\prime\prime}>0$ should  still hold in the right side of $y_s$,
 i.e. $y>y_s$. However, in the right
 side of $y_s$, $g-g^{\prime}y<0$, one has $w_{\phi}<-1$ and $C_{s}^2<0$.
 This is physically forbidden since the k-chameleon becomes unstable.
 As a result, the phase flow cannot pass across $y_s$. Any physically reasonable
 $y$ must be less than $y_s$, which can be seen clearly from Figs.~{2--6} of
  Ref.~\cite{r12} as well. Actually, a general discussion on this point
  has been made in Ref.~\cite{r32}, which shows that a dynamical transition from the
 states with  $w_{\phi}>-1$ to those with $w_{\phi}<-1$ or vice versa is physically
 impossible. One can see that in fact, the condition $y<y_s$ is equivalent to $w_\phi>-1$
 physically.

\subsection{\label{md} Matter-dominated epoch}

In this epoch, the fraction energy density of radiation $\Omega_r\simeq 0$, and
 $\Omega_m+\Omega_\phi=1$ since the universe is spatially flat, as indicated  by
 many astronomical observations~\cite{r4,r5}. In this case, the terms due to the
 strong coupling between k-chameleon and matters cannot be neglected  in the master
 equations Eqs.~{(\ref{eq41})--(\ref{eq44})}. As will be seen
 shortly, due to the appearance of the coupling, the k-chameleon
 model will have a big difference from the ordinary k-essence
 model. In this subsection, we will find out all the possible
 attractor solutions and then study their stability.

\subsubsection{\label{attractorsinmd} Attractors}

At first, we would like to point out that the {\bf D} attractor
[the scalar field tracks the dust ($w_{\phi}(y_d)=0$) during the
matter dominated epoch] and the {\bf S} attractor, which may exist
in the ordinary k-essence model~\cite{r11,r12}, are physically
forbidden in our k-chameleon model, due to the existence of strong
coupling between the chameleon and dust matters (cold dark matter
and baryons). To see this point, let us note that in the cases of
$\Omega_m\simeq 1$, $\Omega_{\phi}\simeq0$ and $w_{tot}\simeq 0$
for the {\bf S} attractor and $w_{tot}=w_{\phi}=0$, but
$\Omega_m\neq 0 $ or $1$ for the {\bf D} attractor, the third term
in Eq.~{(\ref{eq41})} is extremely large (for the case
$\Omega_{\phi} \to 0$) while the first term in Eq.~{(\ref{eq42})}
and Eq.~{(\ref{eq43})} cannot be neglected. Thus, {\em no}
solutions with $y=y_{attractor}=const.$ satisfying $dy/dN=0$ and
$d\Omega_{\phi}/dN=0$, $d\Omega_{m}/dN=0$ exist. Therefore the
attractors {\bf D} and {\bf S} will no longer appear in the
k-chameleon model.

However, we find that the k-chameleon dominated attractor {\bf K}
with $\Omega_{\phi}^{(k)}=1$ can still occur in the k-chameleon
model.  Its characteristics is described by
Eqs.~{(\ref{eq60})--(\ref{eq62}). A remarkable feature we would
like to stress here is that this solution describes a power-law
inflation provided $\sqrt{-g_{k}^{\prime}y_{k}^{2}/2}>1$. As we
will see below, if this attractor is stable, this condition for
accelerated expansion can be satisfied automatically.

Except for the {\bf K} attractor, we find that there is a new
attractor solution in the k-chameleon model, which arises due to
the strong coupling between the k-chameleon field and the dust
matter.  The new attractor is dubbed as {\bf C} attractor (the
letter {\bf C} stands for ``Chameleon"), and has some very
interesting features.  Let us describe the {\bf C} attractor in
some detail.

Considering $\rho_\phi=-Kg^{\prime}$ and $K(\phi)=M^{2}/\phi^2$
[see Eqs.~{(\ref{eq35}) and (\ref{eq40})}] and setting
$M=\sqrt{3}M_{pl}$, from the Friedmann equation Eq.~{(\ref{eq29})}
one has \be{eq64}
H=\frac{\sqrt{-g^\prime}}{\sqrt{\Omega_\phi}}\,\frac{1}{\phi}.
 \ee
Note that the relation $\alpha=\beta/\phi$, we have
 \be{eq65}
\frac{\alpha}{H}=\frac{\beta}{\sqrt{-g^\prime}}\sqrt{\Omega_\phi},
\ee
 which is independent of $\phi$. Therefore it is possible to find an
attractor solution {\bf C} with $y=y_c=const.$ and fixed
$\Omega_{\phi}^{(c)}$, $\Omega_{m}^{(c)}=1-\Omega_{\phi}^{(c)}$
and satisfying the master equations
Eqs.~{(\ref{eq41})--(\ref{eq44})},
 \be{eq66}
r(y_c)=\frac{2+\beta}{2}\sqrt{\Omega_{\phi}^{(c)}}-\frac{\beta}{2}\frac{1}{\sqrt{\Omega_{\phi}^{(c)}}},
\ee and \be{eq67}
\left(\frac{\alpha}{H}\right)_{c}\dot{\phi_c}+3w_{\phi}(y_c)\Omega_{\phi}^{(c)}=0.
\ee
Substituting Eqs.~(\ref{eq65}), (\ref{eq36}) and (\ref{eq23})
into Eq.~(\ref{eq67}), we get
\be{eq68}
\sqrt{\Omega_{\phi}^{(c)}}=-\frac{\sqrt{2}\,\beta}{3g_c}\sqrt{-g_{c}^{\prime}},
\ee
 which is determined only by the coupling constant $\beta$ and the
 function $g(y)$ at $y=y_c$. Substituting $\Omega_{\phi}^{(c)}$ into Eq.~(\ref{eq66}),
 we have
 \be{eq69}
r(y_c)=\frac{9g_{c}^2+2\beta(2+\beta)g_{c}^{\prime}}{6\sqrt{2}\,
g_{c}\,\sqrt{-g_{c}^{\prime}}}.
 \ee
  Comparing with
$r(y_c)=Ms(y_c)=\sqrt{3}M_{pl}\, s(y_c)$ and using
Eq.~(\ref{eq38}), we obtain
 \be{eq70}
g(y_c)=g_c=-\frac{2\beta(2+\beta)}{9y_c}.
 \ee
 Geometrically, this
means that the {\bf C} attractor locates at the intersection of
curve $g(y)$ and the hyperbola
 $h(y)\equiv\disp-2\beta(2+\beta)/(9y)$ in the plot of $g(y)$ versus $y$.
 Because the asymptotes of
 the hyperbola $h(y)$ are the
 two coordinate axes, and $g(y)<0$ when $y>y_d$, this intersection always exists
 in the regime $y_c>y_d$ so  that the k-chameleon contributes a negative
 pressure.
 On the other hand, note that the curve $g(y)$ is monotonically decreasing while the hyperbola
$h(y)$ is monotonically increasing in the regime of $y_d <y <y_s$,
therefore there is only one intersection. In other words, the {\bf
C} attractor always exists and given a function $g(y)$, there is
only one {\bf C} attractor.

Next let us have a look at the other physical features of the {\bf
C} attractor. From Eqs.~(\ref{eq68})
 and (\ref{eq70}), the fraction energy density of the k-chameleon is
\be{eq71}
\Omega_{\phi}^{(c)}=\frac{9y_{c}^{2}(-g_{c}^{\prime})}{2(2+\beta)^2}.
\ee
 The usual  restriction on $\Omega_{\phi}^{(c)}$ is
$0<\Omega_{\phi}^{(c)}<1$. However, as mentioned at the end of
Sec.~\ref{remarks}, the condition $y_c<y_s$ has to be imposed.
 In that case, one has $g_c-g_{c}^{\prime}y_c>0$ while $g_c<0$,
$g^{\prime}_c<0$ and $w_{\phi}(y_c)>-1$. By using
Eqs.~(\ref{eq68}) and (\ref{eq70}), we find that the bounds
 of $\Omega_{\phi}^{(c)}$ should be
\be{eq72} \frac{\beta}{2+\beta}<\Omega_{\phi}^{(c)}<1.
\ee
 Note that $\beta\sim{\cal O}(1)$, this result is quite interesting. For
instance, the lower bound of
 $\Omega_{\phi}^{(c)}$ are $1/3$, $1/2$, $3/5$ and $2/3$ for $\beta=1$, 2, 3 and 4, respectively.
 Therefore, in the {\bf C} attractor, it is not strange that the fraction energy densities of
  k-chameleon and dust matters are comparable. In addition, from Eqs.~(\ref{eq36}), (\ref{eq70}),
  and (\ref{eq71}), one has
\be{eq73}
w_{\phi}(y_c)=-\frac{\beta}{2+\beta}\frac{1}{\Omega_{\phi}^{(c)}}.
\ee
 Thus, the requirement $w_{\phi}(y_c)>-1$ leads to the same
lower bound to $\Omega_{\phi}^{(c)}$ as
 given in Eq.~(\ref{eq72}). Furthermore, from Eqs.~(\ref{eq44}) and (\ref{eq73}), we find
\be{eq74}
w_{tot}(y_c)=w_{\phi}(y_c)\Omega_{\phi}^{(c)}=-\frac{\beta}{2+\beta}.
\ee From the Einstein equation \be{eq75}
\frac{\ddot{a}}{a}=-\frac{1}{6M_{pl}^2}\rho_{tot}(1+3w_{tot}),
\ee
 the universe undergoes an accelerated expansion provided
$\beta>1$, which is equivalent to  $w_{tot}(y_c)<-1/3$. We can
look at this from another angle. Because
$\dot{\phi}_c=\sqrt{2X_c}=\sqrt{2}/y_c$ is constant, one has
 $\phi\propto t$. Then, from Eq.~(\ref{eq64}), $H\propto t^{-1}$ and the scale factor
 $a\propto t^\lambda$. If $\lambda>1$, the universe undergoes a power-law inflation.
 Let us find out the explicit expression of $\lambda$ for the {\bf C}
 attractor. From Eq.~(\ref{eq25}), one has
\be{eq76} \rho_{tot}(y_c)\propto
a^{-3[1+w_{tot}(y_c)]}=a^{-6/(\beta+2)}. \ee Since
$\Omega_{\phi}^{(c)}$ and $\Omega_{m}^{(c)}=1-\Omega_{\phi}^{(c)}$
are both fixed, $\rho_\phi(y_c)$ and $\rho_m(y_c)$ decrease in the
same manner as $\rho_{tot}(y_c) \propto a^{-6/(\beta+2)}$.
 Substituting $a\propto t^\lambda$ and Eq.~(\ref{eq76}) into Eq.~(\ref{eq75}),
  we get $\lambda=(\beta+2)/3$
  by comparing the power of $t$. In fact, from Eqs.~(\ref{eq64}), (\ref{eq71})
  one can find the same result
\be{eq77}
 a\propto t^{(\beta+2)/3}
  \ee
  by using $\phi=\dot{\phi}_c t$. In short, if $\beta>1$, the {\bf
C} attractor solution describes a power-law inflation.

In summary, in the k-chameleon model there may exist  two
attractor solutions, {\bf K} and {\bf C}, and
  {\em no} {\bf D} and {\bf S} attractor solutions in the matter-dominated epoch.
   If $r^2(y)<1$ for any $y<y_s$, the
 {\bf K} attractor cannot exist. Thus, the combination of cosmologically
  relevant attractors during the
 matter-dominated epoch can be one of two types:
\begin{itemize}
\item Only {\bf C} and {\em no} {\bf K};
\item {\bf C} and {\bf K}.
\end{itemize}

\subsubsection{Stability analysis of the attractors}

As mentioned in the beginning of Sec.~\ref{remarks}, the stability analysis
of the attractors {\bf K} and
{\bf C} ought to be treated separately. We study the behavior of small
 deviations from the {\bf K} and
{\bf C} attractor solutions one by one.

(i) {\bf K} attractor. In this case, $\Omega_{\phi}^{(k)}=1$,
$\Omega_{m}^{(k)}=1-\Omega_{\phi}^{(k)}=0$.
 Substituting $y(N)=y_k+\delta y$ and $\Omega_{\phi}(N)=\Omega_{\phi}^{(k)}+\delta\Omega_{\phi}$
 into Eqs.~(\ref{eq41})--(\ref{eq43}) and linearizing these equations, we get
\bea
 &\disp\frac{d\delta
y}{dN}=\frac{3}{2}\,\frac{w_\phi(y_k)-1}{r^{\prime}_k}\left
(r^{\prime}_k\delta y
-\frac{1}{2}\delta\Omega_{\phi}\right ),\nonumber\\
&\disp\frac{d\delta\Omega_{\phi}}{dN}=\left
 (3w_\phi(y_k)+\frac{\sqrt{2}\beta}{\sqrt{-g_{k}^\prime y_k^2}}
\right ) \delta\Omega_{\phi}.\label{eq78}
\eea
 Considering Eq.~(\ref{eq36}), one has $w_\phi(y_k)<0$ since $-g_{k}^\prime>0$
and $g_k<0$. Thus, the solutions of $\delta y$ and
$\delta\Omega_{\phi}$ decay only if \be{eq79}
3w_\phi(y_k)+\frac{\sqrt{2}\beta}{\sqrt{-g_{k}^\prime y_k^2} }<0.
\ee Substituting Eq.~(\ref{eq61}) into it, the stability condition
for the {\bf K} attractor becomes \be{eq80}
\sqrt{-g_{k}^{\prime}y_{k}^{2}/2}>\frac{2+\beta}{3}. \ee Note
that, if $\beta>1$ (the same requirement for the {\bf C} attractor
describes power-law inflation),
 Eq.~(\ref{eq80}) becomes $\sqrt{-g_{k}^{\prime}y_{k}^{2}/2}>1$ which ensures the {\bf K} attractor
 describes power-law inflation too.

(ii) {\bf C} attractor. In this case,
$\Omega_{m}^{(c)}=1-\Omega_{\phi}^{(c)}$. Substituting
$y(N)=y_c+\delta y$ and
$\Omega_{\phi}(N)=\Omega_{\phi}^{(c)}+\delta\Omega_{\phi}$ into
Eqs.~(\ref{eq41})--(\ref{eq43}) and linearizing these equations,
we obtain
 \bea
&\disp\frac{d\delta y}{dN}=B_{1}\delta y+B_{2}\delta\Omega_{\phi},\nonumber\\
&\disp\frac{d\delta\Omega_{\phi}}{dN}=B_{3}\delta y+B_{4}\delta\Omega_{\phi},\label{eq81}
\eea
where
\bea
&\disp B_1=\frac{3}{2}\left[w_{\phi}(y_c)-1\right],\nonumber\\
&\disp B_2=\frac{3}{2}\,\frac{w_{\phi}(y_c)-1}{r^{\prime}(y_c)}\left(-\frac{1}{4\sqrt{\Omega_{\phi}^{(c)}}}\right)
\left(\beta+2+\frac{\beta}{\Omega_{\phi}^{(c)}}\right),\nonumber\\
&\disp B_3=3\Omega_{\phi}^{(c)}\left(\Omega_{\phi}^{(c)}-1\right)w_{\phi}^{\prime}(y_c)+\frac{\beta}{\sqrt{2}}
\frac{g_{c}^{\prime\prime}}{y_c(\sqrt{-g_{c}^{\prime}})^3}\sqrt{\Omega_{\phi}^{(c)}}\left(\Omega_{\phi}^{(c)}
-1\right)+\frac{\sqrt{2}\beta}{y_{c}^2\sqrt{-g_{c}^{\prime}}}\sqrt{\Omega_{\phi}^{(c)}}\left(1-\Omega_{\phi}^{(c)}
\right),\nonumber\\
&\disp B_4=3w_{\phi}(y_c)\left(2\Omega_{\phi}^{(c)}-1\right)+\frac{\beta\left(3\Omega_{\phi}^{(c)}-1\right)}
{y_c\sqrt{2\Omega_{\phi}^{(c)}(-g_{c}^{\prime})}}.\label{eq82}
\eea
From Eq.~(\ref{eq81}), we have
\bea
\frac{d^2\delta y}{dN^2}-(B_1+B_4)\frac{d\delta y}{dN}+(B_{1}B_4-B_{2}B_3)\delta y=0,\nonumber\\
\frac{d^2\delta\Omega_{\phi}}{dN^2}-(B_1+B_4)\frac{d\delta\Omega_{\phi}}{dN}+(B_{1}B_4-B_{2}B_3)
\delta\Omega_{\phi}=0.\label{eq83}
 \eea
 We can see that the solutions of $\delta
y$ and $\delta\Omega_{\phi}$ decay only if
\be{eq84}
B_1+B_4<0~~~~~~~{\rm and}~~~~~~~B_{1}B_4-B_{2}B_3>0.
 \ee
  By using Eqs.~(\ref{eq36}), (\ref{eq38}), (\ref{eq70}), (\ref{eq71}), (\ref{eq73})
  and (\ref{eq82}), and considering $r(y)=Ms(y)=\sqrt{3}M_{pl}s(y)$,
  we can obatin
$$B_1=-\frac{4\beta+2\beta^2+9(-g_{c}^{\prime})y_{c}^2}{6(-g_{c}^{\prime})y_{c}^2},$$
$$B_2=-\frac{(2+\beta)^{2}\left[4\beta+2\beta^2+9(-g_{c}^{\prime})y_{c}^2\right]}
{27g_{c}^{\prime\prime}(-g_{c}^{\prime})y_{c}^4},$$
$$B_3=\frac{3\left[9(-g_{c}^{\prime})^2 y_{c}+\beta(\beta+2)g_{c}^{\prime\prime}\right]
\left[2(\beta+2)^2-9(-g_{c}^{\prime})y_{c}^2\right]}{4\,(2+\beta)^4 (-g_{c}^{\prime})},$$
$$B_4=\frac{\beta\left[2(\beta+2)^2-9(-g_{c}^{\prime})y_{c}^2\right]}{6(2+\beta)(-g_{c}^{\prime})y_{c}^2}.$$
From these quantities, we have
 \be{eq85}
B_1+B_4=-\frac{3(1+\beta)}{2+\beta},
\ee
 \be{eq86}
 B_{1}B_4-B_{2}B_3=\frac{\left[9(-g_{c}^{\prime})y_{c}^2+2\beta(\beta+2)
 \right]\left[2(\beta+2)^2
-9(-g_{c}^{\prime})y_{c}^2\right]}{4(2+\beta)^2
g_{c}^{\prime\prime}y_{c}^3}.
 \ee
Note that  $\beta>0$, $-g_{c}^{\prime}>0$,
$g_{c}^{\prime\prime}>0$ [see Eq.~(\ref{eq39})], and
$\Omega_{\phi}^{(c)}<1$, which implies
$2(\beta+2)^2-9(-g_{c}^{\prime})y_{c}^2>0$ [see Eq.~(\ref{eq71})].
Then it is easy to see that the stability conditions
Eq.~(\ref{eq84}) for the {\bf C} attractor are satisfied. Thus, we
conclude that the {\bf C} attractor is stable.


\subsection{\label{evolution} Cosmological evolution of k-chameleon and
the coincidence problem}

 As mentioned in Sec.~\ref{rd}, in the
radiation-dominated epoch, the behavior of k-chameleon field is
the completely same as that of  the ordinary k-essence without
interaction between the scalar field and background matters
~\cite{r12}. As a result,  three
 kinds of attractors, namely, {\bf R}, {\bf K} and {\bf S}, may exist.
 However, the radiation tracker, {\bf R}, has been argued to has the largest basin of
 attraction on the whole phase plane so that most initial conditions join onto it and
 then makes this scenario became insensitive to initial conditions~\cite{r12}.
  Yet the contribution of
  k-chameleon to the total energy density must not spoil the BBN or not
  dominate over the matter density at the end of the radiation-dominated epoch.
  It has been argued that
  if the contribution of the scalar field energy density in the {\bf R} attractor
  satisfies~\cite{r12}
\be{eq87}
\Omega_{\phi}^{(r)}\simeq 10^{-2}-10^{-1},
 \ee
 the scalar field (k-essence or k-chameleon) will not violate the constraint from the BBN.

The new features appear in the matter-dominated epoch. As
mentioned in Sec.~\ref{md}, in the matter-dominated epoch, the
{\bf D} and  {\bf S} attractors, which may occur in the ordinary
k-essence model,  are physically forbidden in the k-chameleon
model, due to the strong coupling between k-chameleon and dust
matters. Actually, in the k-chameleon model, two kinds of
attractors may exist: one is the familiar {\bf K} attractor and
the other is a completely {\em new} attractor {\bf C}. The new
attractor {\bf C} has some desirable features which can make the
fraction energy densities of k-chameleon and dust matters (cold
dark matter and baryon) be comparable. During the matter dominated
epoch, the relevant attractors can appear in the following two
possible sets: (1)~only {\bf C} and (2)~{\bf C} and {\bf K}. If
$\beta>1$, the universe will undergo a power-law inflation,
regardless the k-chameleon enters into the {\bf C} or {\bf K}
attractor. Suppose that during the radiation dominated epoch, the
universe enters into the {\bf R} attractor, after the onset of
matter domination, it will enter into the {\bf C} attractor with a
large possibility since the values of $\Omega_{\phi}^{(r)}$ and
$y_r$ are required to be somewhat small in order to satisfy the
BBN constraint. In particular, one can adjust the model so that
the {\bf K} attractor does not exist (for instance if $r^2(y)<1$
for $y<y_s$), thus the universe has to enter into the unique {\bf
C} attractor where the fraction energy densities of the
k-chameleon $\Omega_{\phi}$ and the matters $\Omega_m$ are fixed
and keep comparable forever. In this way the k-chameleon model
leads to a natural solution to the cosmological coincidence
problem.

The evolution of the k-chameleon heavily depends on the function
$g(y)$ and other components of the universe like radiation and
dust matter. In Fig.~2 we plot a sketch of possible phase diagrams
of the evolution of the k-chameleon, where only two attractors
${\bf R}$ and ${\bf C}$ appear during the evolution of the
universe. In this plot, we expect that during the radiation
domination epoch, for most initial conditions, the k-chameleon is
attracted to the ${\bf R}$ attractor satisfying the constraint
(\ref{eq87}). In this epoch, the k-chameleon mimics the equation
of state of the radiation component of the universe. With the
increase of $\Omega_m$ for the dust matter component, the
k-chameleon will no longer track the radiation component due to
the interaction between the k-chameleon and dust matter. During
the matter domination epoch, the $\Omega_{\phi}$ will continue to
decrease until $y$ reaches to $y_c$, which can be seen from
(\ref{eq42}) and (\ref{eq43}). Beyond $y_c$, $\Omega_{\phi}$ will
increase and $y$ decreases toward to $y_c$ and passes through it,
which can be seen from (\ref{eq41}). When $y$ decreases to some
value $(<y_d)$, it increases towards to $y_c$, again. After
several such processes, finally the k-chameleon is expected to
reach the stable attractor ${\bf C}$, where the fraction energy
densities of dust matter and dark energies are comparable and the
universe undergoes an accelerated expansion.

 \begin{figure}[ht]
\includegraphics[totalheight=3.5in,angle=90]{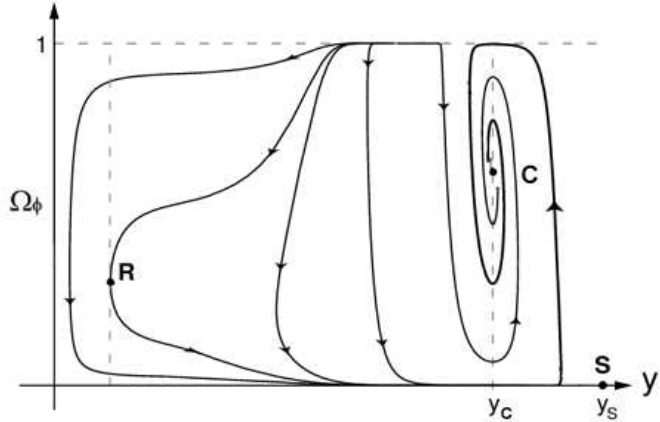}
 \caption{A sketch of possible phase diagrams for the case where only the {\bf R}
 and {\bf C} attractors appear.}
\end{figure}

In order to have a better picture of the k-chameleon model, it is
helpful to do some numerical analysis. The astronomical
observations, such as SNe Ia~\cite{r1,r2,r3}, WMAP~\cite{r4},
suggest that $\Omega_\phi\simeq 0.7$, $\Omega_m\simeq 0.3$, and
$w_\phi<-0.76$ at $95\%$~C.L. today.
 If  adopt $\Omega_\phi(y_c)=0.7$ and impose the constraint $w_\phi(y_c)<-0.75$,
 we see from Eqs.~(\ref{eq72}) and (\ref{eq73}) that
 $2.2<\beta<4.7$ has to be obeyed. From Eqs.~(\ref{eq73}) and (\ref{eq74}), we find
 that $w_\phi(y_c)\simeq-0.86$ and $w_{tot}(y_c)=-3/5$
 for $\beta=3$, while $w_\phi(y_c)\simeq-0.95$ and $w_{tot}(y_c)=-2/3$ for $\beta=4$.
 In these cases, the universe undergoes an accelerated expansion as $a\propto t^{5/3}$
 and $t^{2}$ for $\beta=3$ and $4$, respectively.

In the {\bf C} attractor, the kinetic energy term $X_c=1/y_{c}^2$
is fixed at a constant value. It is possible to design the
function $g(y)$ to get a somewhat large $y_c$. For instance, if
$\Omega_\phi(y_c)=0.7$ and $\beta=3$, we have from
Eq.~(\ref{eq71}) that $y_c\simeq 100$ and $X_c\simeq 10^{-4}$
while $g_{c}^{\prime}\simeq -4\times 10^{-4}$ (the unit
$3M_{pl}^2=1$ has been used here). Comparing $g_{c}^{\prime}\simeq
-4\times 10^{-4}$ with a particular
 example $g_{d}^{\prime}\simeq -5\times 10^{-3}$ in Ref.~\cite{r12}, one can see that
  this value is reasonable, since $y_c\simeq 100\gg y_d=17$.
  Therefore, in this example, the k-chameleon indeed can be treated as a canonical
  chameleon
 approximately. As a result, we cannot detect it from the tests of EP violations
 and fifth force searches on
    the earth or in the solar system today, although it is strongly coupled to
    ambient matters, as illustrated
    in Sec.~\ref{newchameleon}.

 Note that in order to exclude the {\bf K} attractor, one has to
 adjust the model so that $r^2(y) <1$  and decreases monotonically
  in the region $y_r <y <y_s$~\cite{r12}. On the other hand, one
  requires that the {\bf C} attractor exists in the region $y_d
  <y<y_s$. One may wonder whether or not these two conditions can be met
simultaneously. To see this, let us take an example. We have from
Eq.~(\ref{eq66}) that $ r^2(y_c)= [(2+\beta)
\Omega_{\phi}^{(c)}-\beta]^2/(4\Omega_{\phi}^{(c)})$. If $\beta=4$
and $\Omega_{\phi}^{(c)}=0.7$, one then has $r^2(y_c) \simeq
1.4\%$. In the radiation dominated epoch, in order to satisfy the
constraint from the BBN, $ r^2(y_c) = \Omega_{\phi}^{(r)}$ should
be in the region $1-10\%$. So we see that those two conditions can
be satisfied, if $r^2(y_r)> 1.4\%$ and the decreasing of $r^2(y)$
is sufficiently slow so that $y_c$ and $y_s$ can be somewhat large
values. Note that $y_c$ is always less than $y_s$ since $y_s$
locates at $r^2(y_s)=0$.


\section{\label{sect5} CONCLUSION}

Recently a chameleon mechanism has been
suggested~\cite{r20,r21,r22,r23,r24}, in which a scalar field
(chameleon) can be strongly coupled to ambient matters, but it
still satisfies the constraints from the fifth force and EP
violation experiments on the earth and in the solar system. In
this paper we have combined the chameleon mechanism to the
k-essence model of dark energy and have presented a k-chameleon
model. During the radiation dominated epoch, the evolution of the
k-chameleon is the same as that of the ordinary k-essence, and
three kinds of attractors, {\bf R}, {\bf K} and {\bf S}, may
appear. One can construct a model where the {\bf R} attractor has
the biggest basin of attraction so that for most initial
conditions, the universe will be attracted to the {\bf R}
attractor. During the matter dominated epoch, the {\bf D} and {\bf
S} attractors, which may appear in the ordinary k-essence model,
are forbidden in the k-chameleon model, due to the strong coupling
between the k-chameleon and background matters (cold dark matter
and baryons). Except for the familiar {\bf K} attractor, a new
attractor, dubbed {\bf C} attractor, exists in the k-chameleon
model. In the {\bf C} attractor, the fraction energy densities of
the chameleon field (dark energy) and the dust matter (cold dark
matters and baryons) are fixed and comparable, and the universe
enters into an accelerated expansion phase in the power-law manner
if the coupling constant $\beta>1$. We can adjust the model so
that the {\bf K} attractor does not exist, the universe then has
to enter into the {\bf C} attractor. Thus the k-chameleon model
provides a natural solution to the cosmological coincidence
problem.

\section*{ACKNOWLEDGMENTS}
HW is grateful to Ding-Fang Zeng, Xun Su, Fei Wang, Fu-Rong Yin,
and Wei-Shui Xu for helpful discussions. RGC would like to express
his gratitude to the Physics Department, Baylor University for its
hospitality to him during his visit. This work was supported by a
grant from Chinese Academy of Sciences,  grants from NSFC, China
(No. 10325525 and No. 90403029),
 and a grant from the Ministry of Science and Technology of China (No. TG1999075401).


\end{document}